\documentclass[twocolumn, trackchanges]{aastex631}

\newcommand{\scinum}[2]{\ensuremath{#1\!\times\!10^{#2}}}
\newcommand{\figref}[1]{\figurename~\ref{#1}}
\newcommand{\tabref}[1]{\tablename~\ref{#1}}
\newcommand{\secref}[1]{Section \ref{#1}}

\usepackage{amsmath}
\usepackage{layouts}
\usepackage{booktabs}

\begin{document}

\title{Recovering Pulsar Braking Index from a Population of Millisecond Pulsars}

\author[0000-0002-1255-3492]{A. L. Hewitt}
\affiliation{Lancaster University, Physics Building, Lancaster, LA1 4YB, United Kingdom} 

\author[0000-0003-4548-526X]{M. Pitkin}
\affiliation{CEDAR Audio Ltd, Fulbourn, Cambridge, CB21 5BS, United Kingdom}
\affiliation{School of Physics \& Astronomy, University of Glasgow, Glasgow, G12 8QQ, United Kingdom}

\author[0000-0002-2960-978X]{I. M. Hook}
\affiliation{Lancaster University, Physics Building, Lancaster, LA1 4YB, United Kingdom} 

\begin{abstract}

The braking index, $n$, of a pulsar is a measure of its angular momentum loss and the value it takes corresponds to different spin-down mechanisms. For a pulsar spinning down due to gravitational wave emission from the principal mass quadrupole mode alone, the braking index would equal exactly 5. Unfortunately, for millisecond pulsars, it can be hard to measure observationally due to the extremely small second time derivative of the rotation frequency, $\Ddot{f}$. This paper aims to examine whether it could be possible to extract the distribution of $n$ for a whole population of pulsars rather than measuring the values individually. We use simulated data with an injected $n=5$ signal for 47 millisecond pulsars and extract the distribution using hierarchical Bayesian inference methods. We find that while possible, observation times of over 20 years and RMS noise of the order of $10^{-5}$ ms are needed, which can be compared to the mean noise value of \scinum{3}{-4} ms for the recent wideband 12.5-year NANOGrav sample \citep{nanograv_wide}, which provided the pulsar timing data used in this paper.

\end{abstract}

\section{Introduction} \label{sec:intro}

Pulsars are objects of great interest in the world of gravitational waves (GWs). As their pulses are very regular, they can be used as cosmic clocks, and by looking for small, coherent variations over whole populations of pulsars in pulsar timing arrays (PTAs), they have been used to identify the stochastic GW background. This discovery of this background was reported by the North American Nanohertz Observatory for Gravitational Waves (NANOGrav) \citep{2023ApJ...951L...8A}, a joint European–Indian effort with both the European Pulsar Timing Array (EPTA) and the Indian Pulsar Timing Array (InPTA) \citep{refId0}, the Chinese PTA \citep{Xu_2023}, and the Parkes Pulsar Timing Array (PPTA) in Australia \citep{2023ApJ...951L...6R}. This discovery allows us to learn about astrophysical source populations, in particular the distribution of supermassive black hole binary systems.

However, as well as detecting GWs, pulsars are predicted to emit them themselves \citep[e.g.,][]{1969ApJ...157.1395O}. The LIGO-Virgo-KAGRA (LVK) collaboration has already detected several transient gravitational-wave signals from the inspiral and subsequent mergers of neutron stars in binary systems with black holes or other neutron stars \citep{abbott2021gwtc2, 2021ApJ...915L...5A, 2023PhRvX..13d1039A}. While observing transient GWs has become commonplace, it is not the only form of GW expected to be produced by neutron stars. Another category of GWs which remain unobserved to date are continuous waves (CWs), i.e., very long-lived quasi-monochromatic signals. A promising candidate is an individual neutron star spinning with some nonaxisymmetric deformation. Such a deformation could take the form of a mountain on the crust caused by cooling \citep{2000MNRAS.319..902U}, binary accretion \citep{2021MNRAS.507..116G} or due to strong magnetic fields \citep{1996AA...312..675B, 2002PhRvD..66h4025C}. Oscillations beneath the crust, such as $r$-modes, can also cause GW emission from a single pulsar \citep{1998ApJ...502..708A, 1998ApJ...502..714F}. Although these signals are currently yet to be observed, it is possible to obtain evidence of their existence observationally without requiring a GW detection. One method is to measure the braking index $n$ of a pulsar, which we will discuss in this paper.

The braking index $n$ is a measure of a pulsar's angular momentum loss:
\begin{equation}
    \Dot{f} \propto - f^n,
\label{eqn:braking1}
\end{equation}
where $f$ is the pulsar's rotation frequency and $\Dot{f}$ is the first time derivative of the frequency, i.e., the spin-down rate. The value of $n$ depends on the mechanism of angular momentum loss. For example, a pulsar losing angular momentum through magnetic dipole radiation follows
\begin{equation}
    \Dot{f}_{\rm dip} = - \frac{8\pi^2}{3c^3}\frac{B^2}{I_{zz}}R^6f^3\sin^2{\alpha},
\end{equation}
which gives a braking index of 3 \citep[e.g.,][]{Hamil_2015}. Here, $c$ is the speed of light, $B$ represents the magnitude of the dipolar magnetic field at the magnetic pole, $I_{zz}$ is the moment of inertia, $\alpha$ is the angle between the magnetic field axis and the rotation axis, and $R$ is the star's radius at the magnetic pole. 

Similarly, if the spin-down of a non-axisymmetric pulsar was instead dominated by emission of gravitational waves (as in the case of a `gravitar') the angular momentum loss would follow 
\begin{equation}
    \Dot{f}_{\rm GW} = -\frac{512\pi^4}{5}\frac{G}{c^5}I_{zz}\varepsilon^2f^5,
\end{equation}
where $\varepsilon \equiv |I_{xx}-I_{yy}|/I_{zz}$ is the ellipticity of the pulsar and $(I_{xx}, I_{yy}, I_{zz})$ are the source's principle moments of inertia \citep{2005palomba}. Here $G$ represents the gravitational constant. In this case, the braking index is equal to 5. \cite{min_ellipticity} observed a possible minimum ellipticity for millisecond pulsars (MSPs) consistent with $n=5$, suggesting that the spin-down of pulsars near this limit is dominated by GW emission. It is also possible for pulsars to have a braking index of 7 if gravitational waves are emitted via $r$-modes \citep{2021ApJ...922...71A}. Most realistic systems are expected to have a mixture of several processes such that $\dot{f}=-af^3-bf^5$, where $a$ and $b$ are coefficients that define the amount of mixing.

By solving the differential equation in equation \eqref{eqn:braking1}, the value of $n$ can be determined from observational parameters using 
\begin{equation}
    n = \frac{f\Ddot{f}}{\Dot{f}^2}.
\label{eq:n}
\end{equation}
This requires knowledge of both the first and second derivatives of $f$. Although several pulsars have measured braking indices \citep{obs_n, glitches}, the second derivative for MSPs is usually small, often around the order of $10^{-30}$\,s$^{-3}$, which is difficult to observe over current timescales.
 
Rather than trying to directly measure the braking index of single pulsars, this paper discusses the possibility of using a population of pulsars to infer the distribution of $n$ for the whole sample. This could allow the underlying energy loss process to be inferred without needing precise values of $\Ddot{f}$ for individual sources. If the braking index distribution could be determined in this way, it would give valuable insight into whether pulsar populations are spinning down due to GW emission. 

In this paper, we discuss the method used to obtain and extract the braking index from the pulsar population in \secref{sec:analysis}, present and discuss the results in \secref{sec:results}, and comment on the feasibility of this method now and for future observations in \secref{sec:discussion}.

\section{Analysis}
\label{sec:analysis}

We create simulated time of arrival data (TOAs) for simplified pulsar models with injected braking indices of primarily $n=5$ (see \secref{sec:vary-n} in the Appendix for analysis using other values of $n$), representing pulsars with spin-down dominated by GW emission. Posteriors on $n$ for each pulsar are then obtained from their TOAs without an informed assumption on the value of $n$. These posteriors are combined to extract the distribution of $n$ for the entire population. We look at how factors like noise levels, observation length, and observation frequency affect the accuracy and confidence of the recovered distribution.

\subsection{Creating TOAs}
\label{sec:make_fake}

A sample of $47$ pulsars from the wideband NANOGrav 12.5-year data set \citep{nanograv_wide, nanograv_narrow} are used as example pulsars for this analysis. All pulsars in this sample are MSPs, meaning they have a rotation frequency of over 100\,Hz. This puts the sample closer to the cut-off described in \cite{min_ellipticity} which is caused by $n=5$ processes than if it contained pulsars with lower frequencies. It may be noted that 30 of the pulsars are in binary systems; however, this information is not used in this study and they are treated as isolated pulsars. 

We take the pulsar parameter files provided by NANOGrav and strip them of unnecessary parameters, leaving a basic imitation of the original pulsar: binary information is removed so all pulsars appear isolated and sudden rotation spin-up events called glitches are ignored. The effect of glitches on the braking index is discussed briefly in \secref{sec:discussion}. The parameters retained are chosen as only those required for the production of the fake TOAs by the pulsar timing software package \texttt{Tempo2} \citep{tempo21, tempo22, tempo23}. A list of the parameters can be found in \tabref{tab:kept_params} in the Appendix. The values \texttt{F0} and \texttt{F1} (representing $f$ and $\dot{f}$) are used as provided, while \texttt{F2} ($\ddot{f}$) is fixed at the value corresponding to a braking index of $n=5$ by rearranging equation \eqref{eq:n}. This provides realistic TOA data for pulsars with an injected $n=5$ for us to test the analysis method on.

Using simulated TOAs instead of real data decreases the number of variables, allowing the analysis to be performed with fewer unknown effects. It also allows us to vary parameters such as noise and frequency of observations. We can investigate beyond current observational constraints, performing the analysis with observation times that exceed the 12.5 years of the NANOGrav sample. Importantly, it allows us to know the true value of $n$ so we can verify the accuracy of our results.

Real pulsar TOAs are subject to noise, which is generally classified into two types: white (uncorrelated in time) and red (correlated in time). Red noise is prevalent at low frequencies and so affects data with longer timescales. There are several sources of red noise: stochastic GWs which will induce correlated red noise in the residuals (the difference between the predicted TOAs and the actual TOAs) of multiple pulsars, processes intrinsic to each pulsar (such as companions), and the varying effects of the interstellar medium (ISM). The spindown of younger pulsars is caused by various complex methods \citep{2000A&A...354..163P}, leading to stronger red noise than their older MSP counterparts. Although sources of red noise can be hard to identify, the noise properties of the MSP pulsars in the NANOGrav sample can mostly be described as white noise \citep{noise2}, with Gaussian noise present at all frequencies. White noise represents instrumental errors that have not been modelled and intrinsic pulse jitter \citep{2011MNRAS.417.2916L, 2021MNRAS.502..407P}. The impact of noise, especially red noise, on high-precision pulsar timing and its effect on the measurability of $\Ddot{f}$, can be seen in \cite{Liu19}.

In this analysis, we allow our simulated TOAs to be created with a root mean square (RMS) noise value for white noise alone. For each pulsar, the noise value is taken from Table 5 in \cite{nanograv_wide}. In cases where the source has a separate red noise value, the value for white noise is taken only. These values are listed in \tabref{tab:pul_vals}. As the effect of timing noise on the accuracy of this method will be tested, we need to be able to vary the average noise of the sample to any desired value $\text{RMS}_{\text{desired}}$. Therefore, we scale the RMS noise of each individual pulsar using
\begin{equation}
\label{eq:rms}
    \text{RMS}_{i, \text{scaled}} = \mu_\text{desired}\times\frac{\text{RMS}_i}{\mu},
\end{equation}
where RMS$_i$ is the RMS noise value for pulsar $i$ taken from \cite{nanograv_wide}, $\mu$ is the mean of the original dataset, and $\mu_\text{desired}$ is the mean the sample will be scaled to. This retains the realistic natural variation of RMS noise throughout the population.

The number of days between observations is kept at 28 throughout this analysis, despite it being one of the parameters which we wish to vary. Instead, we simulate changing it using the known relationship between frequency of observations and RMS noise for a pulsar assuming only white noise:
\begin{equation}\label{eq:cadence}
    \text{RMS}_{\text{eqv}} = \frac{\text{RMS}}{\sqrt{N}},
\end{equation}
where $N$ is the factor by which the cadence of observations is increased and RMS$_\text{eqv}$ is the RMS noise equivalent to such an increase in observations \citep{noise2}. This allows us to simulate a higher cadence without increasing the computational load. The full list of parameters used in generating the fake TOAs are listed in \tabref{tab:run_params} in the Appendix. 

\subsection{Obtaining posteriors}
\label{sec:enterprise}

We choose to use \texttt{enterprise} \citep{enterprise} and \texttt{enterprise\_extensions} \citep{entex} to model TOAs and provide likelihoods using our simulated TOAs for each pulsar\footnote{We have used a modified version of \texttt{enterprise\_extensions} \citep{entex_edit} that retains 128 bit floating point precision for parameter values to avoid numerical truncation errors.}. \texttt{enterprise\_warp} \citep{entwarp}, which is a set of tools which allow Bayesian inference via the \texttt{bilby} package \citep{bilby1, bilby2} in combination with \texttt{enterprise}, is used to produce posteriors on the rotational parameters that we require. \texttt{enterprise\_warp} can only sample over non-derived pulsar parameters, i.e., the rotational frequency and its derivatives. However, the very strong correlation between $\dot{f}$ and $\ddot{f}$ means it is difficult to draw samples from their joint posterior. We have therefore modified \texttt{enterprise\_warp} \citep{enterprise_warp_edit} to allow it to draw samples directly from the braking index so, for each pulsar, we sample from the posterior on $f$ (\texttt{F0}), $\dot{f}$ (\texttt{F1}), $n$ and a constant white noise variance. We use priors on $f$ and $\dot{f}$ defined by the \texttt{enterprise\_extensions} defaults of uniform priors that are constant between $\pm5 \sigma$ (where $\sigma$ is taken from the values from the fit provided in the PTA parameter files). For the braking index we use a uniform prior that is constant between $0 \le n \le 10$. For the sampling we use the \texttt{dynesty} nested sampling package \citep{dynesty}, via \texttt{bilby}. Despite not being required for our subsequent analysis, $\ddot{f}$ samples can be recreated from those we obtain via equation~\eqref{eq:n}.

\subsection{Recovering underlying distribution}
\label{sec:stacker}

Finally, the posteriors on $n$ can be ``stacked'' using the python package \texttt{posteriorstacker} \citep{posteriorstacker, posterior_method} to infer and plot the underlying distribution of $n$ for the entire population.

\texttt{Posteriorstacker} uses the hierarchical Bayesian model described in Appendix A of \cite{posterior_method} to infer the intrinsic distribution for desired parameters given posteriors for a sample of objects. We briefly describe the method here. The parent distribution of a value $n$ for all objects is assumed to be a Gaussian $N(n|\mu, \sigma)$ with unknown mean $\mu$ and standard deviation $\sigma$ (the hierarchical model's {\it hyperparameters}); in this case the objects are pulsars and $n$ is the braking index. 
As the parent distribution should hold for all pulsars, the hierarchical Bayesian modelling likelihood for the hyperparameters can be calculated as the product of the individual pulsar likelihoods marginalised over $n$
\begin{equation}\label{eq:like}
    \mathcal{L} \equiv P(\mathbf{D}|\mu, \sigma) = \prod_i \int P(D_i|n)N(n|\mu, \sigma)\text{d}n.
\end{equation}
where $\mathbf{D}$ refers to the full dataset of all 47 pulsars used in this analysis.
The posterior distributions on the braking index, $P(n|D_i)$, for each pulsar, $i$, can be used as likelihoods in equation \eqref{eq:like}, i.e.,
\begin{equation}
    P(D_i| n) \propto P(n|D_i),
\end{equation}
which is valid due to having used a uniform prior on $n$ in the posterior inference, with the associated constant of proportionality being unimportant. As we have a finite number of posterior samples rather than a functional form of the posterior, the \texttt{posteriorstacker} package makes use of the useful observation that
\begin{equation}
\int P(D_i|n)N(n|\mu, \sigma)\text{d}n \approx \frac{1}{m_i}\sum_{j=1}^{m_i}{N(n_{ij}| \mu, \sigma)},
\end{equation}
where $m_i$ is the number of samples (usually on the order of 20,000 per pulsar) and $n_{ij}$ are the posterior sample values of the braking index for the $i^{\rm th}$ pulsar, i.e., it is the expectation value/mean of the parent distribution evaluated at the posterior sample values. So, the likelihood that \texttt{posteriorstacker} evaluates over $\mu$ and $\sigma$ can be approximated by
\begin{equation}\label{eq:approxlike}
    \mathcal{L} \approx \prod_i \frac{1}{m_i}\sum_{j=1}^{m_i}{N(n_{ij}| \mu, \sigma)}.
\end{equation}

Equation \eqref{eq:like} or \eqref{eq:approxlike}, can be converted to a posterior over $\mu$ and $\sigma$ by multiplying by appropriate priors on these hyperparameters. In this case, the priors are uniform for $\mu$ and log-uniform for $\sigma$. 

\texttt{Posteriorstacker} provides two model distributions: histogram (using a Dirichlet prior distribution) and Gaussian. The histogram distribution is more agnostic of the true underlying distribution, but it is therefore also less constraining. We use the Gaussian distribution as it provides better constraints and has fewer parameters to infer (i.e., it's a ``simpler'' model). For the Gaussian model, \texttt{posteriorstacker} estimates the distributions for the mean and standard deviation for the braking index. \texttt{Posteriorstacker} can provide an evaluation of the parent Gaussian distribution at different percentiles of the sampled hyperparameters, i.e., the median (50th percentile) distribution. The solid lines in Figures \ref{fig:rms}-\ref{fig:num} come from generating Gaussians using the median values from these distributions, while the shaded bands come from the values at the 95th and 5th percentiles of the distributions.

To summarise \secref{sec:analysis} so far, for each analysis presented below, we perform the following steps: i) generate simulated TOAs for a set of pulsars, ii) for each pulsar, use the simulated TOAs to draw posterior samples on $f$, $\dot{f}$ and $n$ using a nested sampling algorithm, and iii) use the combined posterior samples on $n$ from all pulsars to infer posterior samples on the mean $\mu$ and standard deviation $\sigma$ of a parent Gaussian distribution for $n$.

\subsection{Quantifying the results}
\label{sec:quantify}

The results obtained in this analysis are plots, which provide easy visual analysis of the degree to which recovering $n$ is successful. We also want to be able to quantify these results numerically by calculating the odds ratio that a randomly drawn braking index would be within a given range about the desired value. To achieve this, we begin with the following equation
\begin{align}
\label{eq:quantify}
\int_{a}^{b} \frac{\sigma}{\sqrt{2\pi}} \exp\left(-\frac{(x-\mu)^2}{2\sigma^2}\right) dx &= \notag \\
\frac{\sigma^2}{2} \Bigg[ \text{erf}\left(\frac{(\mu-a)}{\sqrt{2}\sigma}\right) &- \text{erf}\left(\frac{(\mu-b)}{\sqrt{2}\sigma}\right) \Bigg]
\end{align}
which is the analytical solution to integrating the Gaussian distribution from $a$ to $b$, where $\mu$ and $\sigma$ are the mean and standard deviation. 

First, we decide on an arbitrary grid on the braking index. Then for each grid bin we evaluate equation \eqref{eq:quantify} using the grid bin boundaries, $b_\text{min}$ and $b_\text{max}$, as the limits of integration for the all the equally weighted posterior sample values distributions. We then sum the probabilities within each bin. This gives the probability, $\text{O}_{b_\text{min}-b_\text{max}}$, that a braking index randomly drawn from our distributions would be within that bin: 
\begin{align}
\label{eq:quantify2}
& \text{O}_{b_\text{min}-b_\text{max}} = \notag \\
& \sum_k^{N_k} \frac{\sigma_k^2}{2} \Bigg[ \text{erf}\left(\frac{(\mu_k - b_\text{max})}{\sqrt{2}\sigma_k}\right) - \text{erf}\left(\frac{(\mu_k-b_\text{min})}{\sqrt{2}\sigma_k}\right) \Bigg],
\end{align}
where $N_k$ is the number of samples from the posterior of the underlying distribution hyper-parameters $\mu$ and $\sigma$.

To find the equivalent probability within an arbitrary range, for example $[4,6]$, we can sum up the probabilities for the bins within that range. We can then also sum together the probabilities for all bins, $\text{O}_{\text{all}}$ (in our case, this is $[0, 10]$) to obtain the odds ratio:
\begin{equation}
    \text{OR}_{a-b} = \frac{\text{O}_{a-b}}{\text{O}_{\text{all}}-\text{O}_{a-b}}.
\end{equation}

\begin{figure}[ht]
    \centering
    \includegraphics[width=\linewidth]{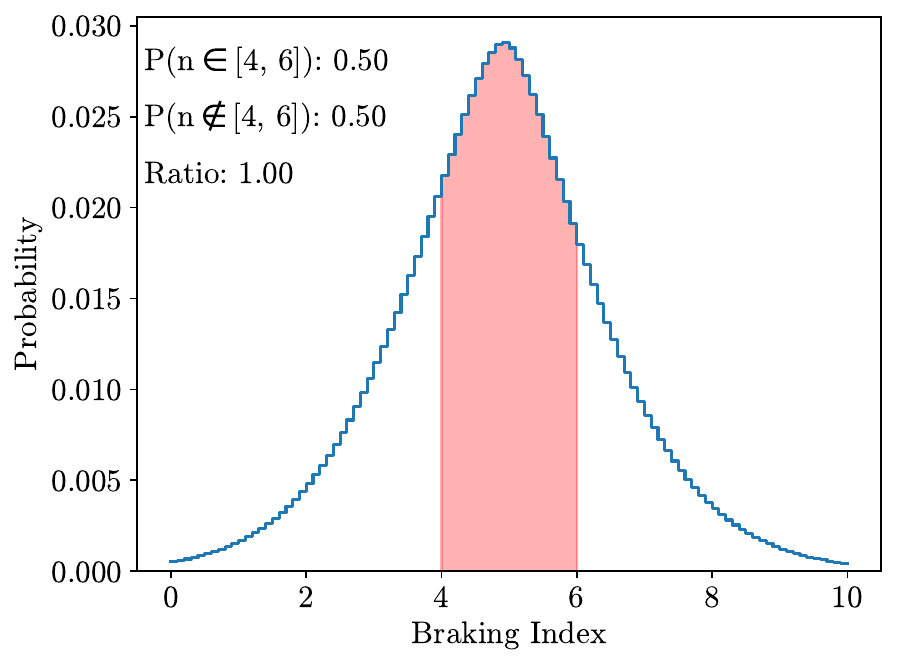}
    \caption{A normalised histogram of the equally weighted posterior samples binned into 100 bins for an observation length of 20 years and an RMS noise value of \scinum{1}{-5} ms using all 47 pulsars. Using equation \eqref{eq:quantify}, probability that $n$ lies in the bins between 4 and 6 (shaded in red) and the probability that it lies outside that region is calculated to find $\text{OR}_{4-6}$.}
    \label{fig:odds_ratio}
\end{figure}

In this analysis, we choose to use 100 bins and calculate the odds ratio within the range $[4,6]$, denoted as $\text{OR}_{4-6}$. While the choice is arbitrary, this range is chosen as it excludes the $n=3$ case where the spin-down is due to magnetic dipole radiation alone and $n=7$ case where it is due to GW emission via r-modes. \figref{fig:odds_ratio} visualises the results of this process for a run with an observation length of 20 years, an RMS noise value of \scinum{1}{-5} ms using all 47 pulsars. For this run, $\text{OR}_{4-6} = 1.00$, meaning there is a 50\% chance of drawing a value of $n$ between 4 and 6 from the distributions.

\section{Analysis results}
\label{sec:results}

Following the methods described in \secref{sec:analysis}, we repeat the analysis several times, changing the RMS noise, observation length, and number of pulsars to identify the parameters with the greatest impact on the recoverability of the braking index using this method. In the following sections, the ``default'' parameters will be an observation length of 20 years with a mean RMS noise of $1\times10^{-5}$ ms using all 47 pulsars. The run using these parameters is depicted as the solid blue line in each plot.

\subsection{RMS noise}

\begin{figure}[ht]
    \centering
    \includegraphics[width=\linewidth]{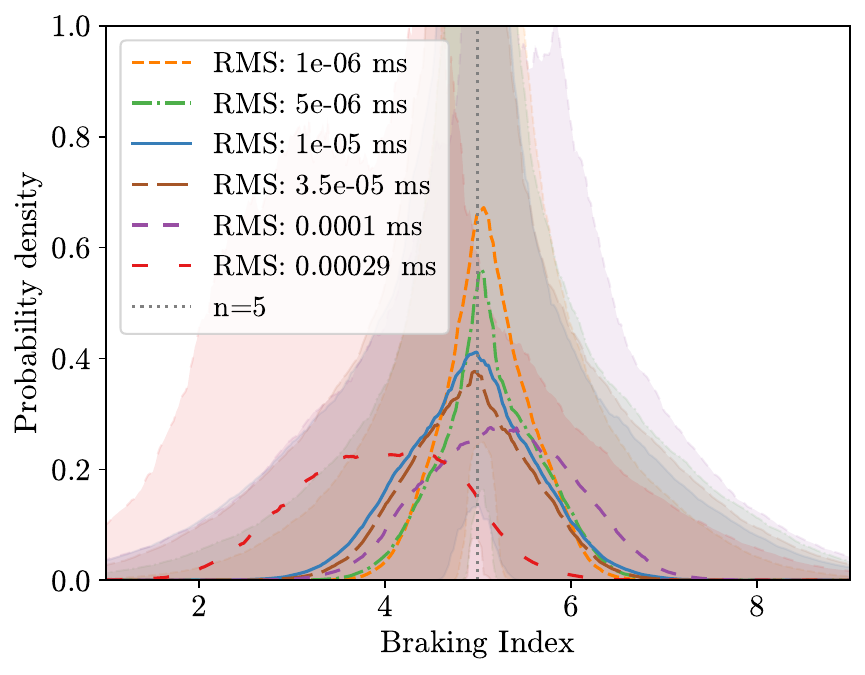}
    \caption{Analyses with identical parameters aside from varying the mean RMS noise, calculated using equation \eqref{eq:rms}, in the simulated TOAs. The lines represent the median values while the shaded regions are bounded by the 5th and 95th percentiles. The $\text{OR}_{4-6}$ for these runs, in order from highest rms to lowest, are: 0.62, 0.88, 0.98, 1.00, 1.06, 1.76.}
    \label{fig:rms}
\end{figure}

Of the three main factors we want to investigate, we look at the RMS noise value first. It makes sense to include the full pulsar population for these comparisons, since pulsars can be analysed in parallel so they don't affect the runtime. Performing this analysis for a 50 year observation length could cost over 262 GB of memory and take on the order of a week to run, so care was taken to avoid running lots of expensive analyses. Therefore, an observation length of 20 years was chosen as a for repeat analyses. This value is also ideal as in a few years the pulsars in this sample will have 20 years of observations. Changing the mean RMS noise has two effects, as discussed in \secref{sec:make_fake}. While it represents white noise on the pulsar TOA data, it can also represent changing the cadence of observations using equation \eqref{eq:cadence}. It is important to note that the cadence of observations is fixed at 1 per month for all analyses. The following values for RMS (in ms) are chosen:
\begin{itemize}
\item \scinum{2.9}{-4} ms: The mean RMS for the NANOGrav sample.
\item \scinum{1}{-4} ms: It is estimated that the Square Kilometer Array (SKA) will produce this RMS for 50 pulsars \citep{ska}.
\item \scinum{3.5}{-5} ms: The smallest RMS for a pulsar in the NANOGrav sample (J2234+0611).
\item \scinum{1}{-5} ms: The SKA could provide this RMS value for five pulsars \citep{ska}. 
\item \scinum{5}{-6} ms: The equivalent RMS corresponding to a true RMS of \scinum{1}{-5} ms but with an observation every week rather than every month.
\item \scinum{1}{-6} ms: Chosen to demonstrate what kinds of rms values would have a strong effect on the confidence.
\end{itemize}

Where possible, one set of TOAs is produced and repeated analyses performed on that set with the parameters for the inference analysis varied as desired. This ensures the least amount of random error is introduced. However, with the RMS and observation time comparisons, new TOA data must to be produced for each run as RMS noise is introduced when producing the simulated data. This means these analyses are subject to additional variation caused by the independent noise realizations. This is likely to be minimal, as discussed in \secref{sec:posteriors} in the Appendix.

The results for the various RMS values can be seen in \figref{fig:rms}. The lines represent the median values of a Gaussian braking index distribution for each RMS value, and the shaded regions are bounded by the 5th and 95th percentiles. The RMS noise values below \scinum{3.5}{-5} ms are able to recover the $n=5$ signal with $\text{OR}_{4-6}$ over 1.00, meaning it is more likely that the braking index is between 4-6 than outside that range (such as 3 or 7 which would be expected from magnetic dipole radiation or r-modes respectively). This suggests reasonable accuracy and confidence. The difference between $\text{OR}_{4-6}$ for the factor of 10 improvement between \scinum{1}{-5} ms and \scinum{1}{-6} is a factor of 1.76. This suggests that for current optimistic projections, decreases to TOA noise will not greatly improve the viability of this analysis method. The RMS noise value of \scinum{1}{-6} was included to verify that the braking index could be recovered with vast improvements in noise, which it clearly shows. Values even smaller than this were looked at, and once the change described in \secref{sec:enterprise} was made to \texttt{enterprise\_extensions}, they continued this trend.

A mean RMS noise value of \scinum{1}{-5} calculated using equation \eqref{eq:rms} was chosen to be kept constant for the subsequent analyses as it represented the most optimistic value which was still realistic. 

\subsection{Observation length}
\label{sec:length}

\begin{figure}[ht]
    \centering
    \includegraphics[width=\linewidth]{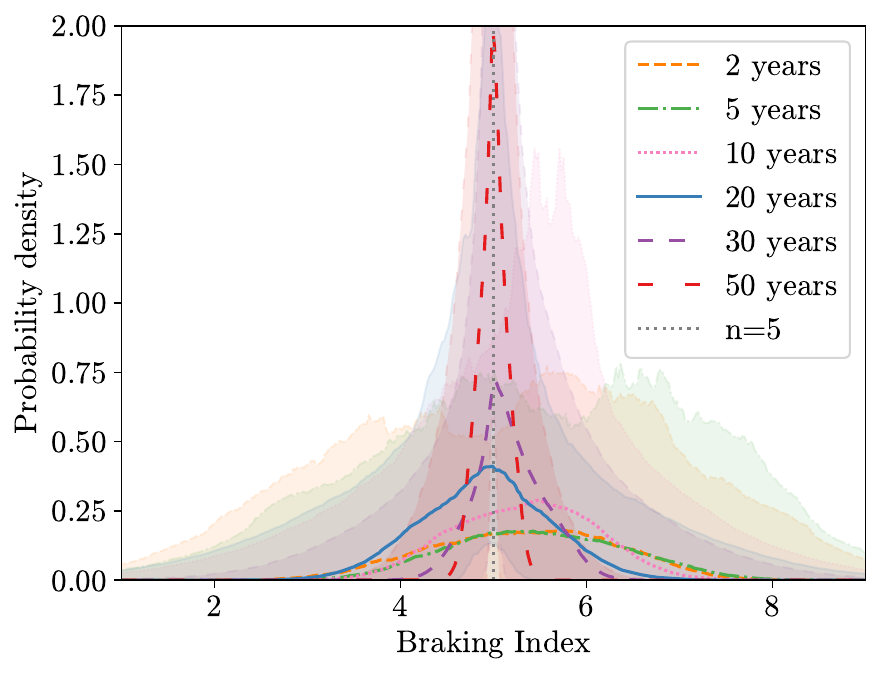}
    \caption{Analyses with identical parameters aside from varying observation time. The solid lines represent the median values while the shaded regions are bounded by the 5th and 95th percentiles. The $\text{OR}_{4-6}$ for these runs, in order from shortest to longest observation length, are: 0.69, 0.69, 0.92, 1.00, 1.46, 9.65.}
    \label{fig:length}
\end{figure}

Since $n$ depends on the frequency double derivative which changes very slowly over time, we expect more accurate and confident results with increased observation time as the change over that period will be larger. This is especially true as $\Ddot{f}$ will have a cubic increase with time. Additionally, $\Ddot{f}$ decorrelates with $\dot{f}$ over time, making it easier to measure. \figref{fig:length} shows the results of five runs of varying observation lengths. In this case, the simulated TOAs are created for each analysis, with different specified lengths going backwards in time. This could theoretically result in slightly better results due to higher pulsar $f$, $\dot{f}$, and $\Ddot{f}$ values than in future measurements. However, this, and the location of the epoch within the observation length, was verified to make negligible difference. The mean RMS noise is kept at \scinum{1}{-5} ms and we include all 47 pulsars, then identical posterior analyses is performed.

It can be seen that the confidence increases with length of observation. For 10 or fewer years, the correct braking index is not recovered with any significant confidence reflected by $\text{OR}_{4-6} <$  1.00. For 20 years, the mean $n$ is correct, but with a $\text{OR}_{4-6}$ value of 1.00 exactly. For longer observation lengths, the correct $n$ was recovered with an $\text{OR}_{4-6}$ of 1.46 and 9.65 for 30 and 50 years, respectively. For the 50 year run, the signal at $n=5$ is very significant, although multiple decades must pass before such an observation length is realistic. That time-frame would doubtlessly also herald advances in timing precision which would also improve the probability of a signal. However, it should be remembered that these runs were performed assuming an already very optimistic RMS noise value.

\subsection{Number of pulsars}
\label{sec:num}

\begin{figure}[ht]
    \centering
    \includegraphics[width=\linewidth]{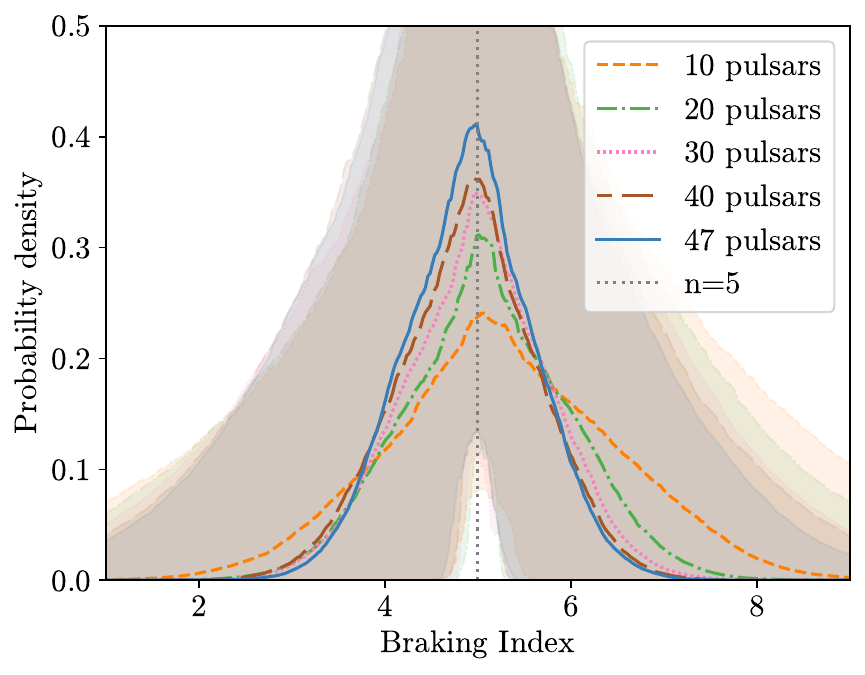}
    \caption{Analyses with identical parameters aside from varying the number of pulsars. For each set, additional pulsars are picked at random with the exception of the pulsar with the strongest signal, which was included in all runs. The solid lines represent the median values while the shaded regions are bounded by the 5th and 95th percentiles. The $\text{OR}_{4-6}$ for these runs, in order from fewest to the most pulsars, are: 0.53, 0.70, 0.81, 0.90, 1.00.}
    \label{fig:num}
\end{figure}

\figref{fig:num} shows the results of running the analysis with different numbers of pulsars. In this case, the same TOA and posterior data is used with an observation time of 20 years and an RMS of \scinum{1}{-5} ms. Originally, the pulsars to be included were chosen randomly, but it was identified that one pulsar with a well-constrained braking index dominates the results, causing dramatic improvements the moment it was included. To limit this effect, the pulsar is automatically included in the dataset for the first 10 pulsars, and all subsequent datasets. 
Reproducing this plot results in variation only due to the random selection of pulsars added to each run. As expected from keeping the data for each pulsar constant, the median does not dramatically change between results.

As can be seen, increasing the number of pulsars improves the confidence of the detection of $n$. While the signal is dominated by one exceptional source which is included from the start, it can be improved by increasing the population size as expected. There is roughly a factor of 2 improvement between the maximum probability densities for the 10 and 47 pulsar runs, which demonstrates the significance of adding more pulsars, even if they are not more well constrained than those already included in the sample.

The pulsar selected to be included in the 10 pulsar run was B1937$+$21 with $n = 5.01\pm0.06$. This pulsar has the highest frequency derivative and therefore also the highest double derivative, as can be seen in \tabref{tab:pul_vals}.

\section{Discussion}
\label{sec:discussion}

This analysis looks at purely simulated data which has been simplified to represent isolated versions of real pulsars. Therefore, when performing this analysis on real data it is reasonable to expect that it would be harder to extract the underlying distribution of $n$. Specifically, the inclusion of parameters which are correlated with $\Ddot{f}$, such as the pulsar's proper motion or non-white timing noise, would make this more complicated. We also assume that the measured $\dot{f}$ is dominated by the intrinsic spin-down and not greatly contaminated by acceleration effects. The results in this paper therefore represent an optimistic scenario, but are presented as a proof-of-principle study.

Of the factors considered, improvement of some are more achievable than others. Although some pulsars have been observed for 50 years, such as Vela, having a large sample with consistent observations and small timing residuals for such a time will not happen soon. For example, the NANOGrav dataset \citep{nanograv_narrow} used in this analysis has 12.5 years of observations for 47 pulsars, while the EPTA has 24 years for 6 pulsars \citep{24years} and the PPTA as observed 30 millisecond pulsars spanning up to 18 years \citep{2023ApJ...951L...6R}. The newest NANOGrav dataset, the 15 year dataset, has 67 pulsars \citep{2023ApJ...951L...8A}. These three groups are part of the International Pulsar Timing Array (IPTA). In the future, combined arrays such as this will allow for much larger populations of pulsars to be analysed for longer periods of time, thus increasing the likelihood that this analysis would observe the underlying distribution on the braking index.

Increasing the cadence of observations is known to have the same effect as reducing the RMS noise. However, even with collaborations such as the IPTA, it may be more beneficial to have individual telescopes observing different pulsars than for multiple telescopes observing the same source \citep{cadence}. Therefore, improving the RMS noise itself would be preferable as it would free observation time for other targets. The SKA, which is currently in construction and expected to enter operation by the end of the decade, could provide an RMS of \scinum{1}{-4} ms to 50 pulsars and potentially \scinum{1}{-5} ms to five more \citep{ska}. 

\section{Conclusions}
\label{sec:conclusions}

This paper looks at the feasibility of observing the braking index, $n$ of a population of pulsars in order to identify the types of processes involved in angular momentum loss. Simulated pulsar TOA data with an injected braking index of $n=5$ is produced using \texttt{Tempo2} \citep{tempo21, tempo22, tempo23}. Without assuming any prior knowledge of $n$, posteriors are produced by \texttt{enterprise} \citep{enterprise} for each pulsar, which are then combined using the hierarchical Bayesian model in \texttt{posteriorstacker} \citep{posteriorstacker}. The effect of RMS noise, observation length, and number of pulsars in the sample are investigated. 

It is found that with realistically optimistic values for RMS noise, we are able to accurately recover the braking index from the sample, but with a low $\text{OR}_{4-6}$ of around 1.0, which means it is equally likely that $n$ is between $4-6$ as it is outside that range. Decreasing the RMS noise to values better than predicted for next-generation detectors enables us to accurately and confidently recover the signal, but order of magnitude improvements would be needed to increase the ratio by a factor of 1.5. Due to the difficulty of such an improvement, this parameter is likely to have the least effect on the feasibility of this method in the near future.

The observation length is found to have a large impact on how well $n$ is recovered with this method. Using an optimistic value for the RMS noise, a 20-year observation length enables accurate identification of $n$ with an $\text{OR}_{4-6}$ of 1.00. For longer observation lengths, the probability increases significantly to 9.65 for 50 years. This parameter has a big impact on the feasibility of this method, but it would require many years to pass.

The number of pulsars is also found to increase the likelihood of confidently recovering a signal significantly, with a factor of four increase in the number of pulsars resulting in a factor of 2 increase in $\text{OR}_{4-6}$. With more telescopes being built and the increase in large collaborations, this parameter should not be too difficult to improve, and therefore is likely to have a large impact on the feasibility of this method. 

In conclusion, we have developed a method for extracting the braking index distribution for a set of millisecond pulsar observations. In our proof-of-principle studies, we have shown that this works for a simplified scenario under some optimistic assumptions. While this may not provide a way to confidently constrain a particular pulsar braking mechanism if applied to current real timing datasets, this may be possible with future observations. It is still worthwhile testing the method on real datasets, with some of our assumptions and simplifications relaxed, and seeing what constraints can be placed on values of the braking index.

\begin{acknowledgments}
We are grateful for computational resources provided by Cardiff University, and funded by STFC awards supporting UK Involvement in the Operation of Advanced LIGO through grants ST/N000064/1 and ST/V001337/1. IH acknowledges support from STFC through grants ST/Y001230/1 and ST/V000713/1.
\end{acknowledgments}

\vspace{5mm}
\facilities{NANOGrav \citep{nanograv_wide, nanograv_narrow}}

\software{\texttt{enterprise} \citep{enterprise},
          \texttt{enterprise\_extensions} \citep{entex, entex_edit},
          \texttt{enterprise\_warp} \citep{entwarp, enterprise_warp_edit}
          \texttt{posteriorstacker} \citep{posteriorstacker}, 
          \texttt{bilby} \citep{bilby1, bilby2, bilby_mcmc}, 
          \texttt{numpy} \citep{numpy}, 
          \texttt{Astropy} \citep{astropy:2013, astropy:2018}, 
          \texttt{Tempo2} \citep{tempo21, tempo22, tempo23},  
          \texttt{matplotlib} \citep{2007CSE.....9...90H}, 
          }

\pagebreak

\appendix

\label{sec:appendix}

\section{Tables of parameters}

\begin{table}[ht]
\centering
\begin{tabular}{ll}
\hline
Parameter & Details \\
\hline
PSR & Pulsar name \\
LAMBDA & Ecliptic longitude \\
BETA & Ecliptic latitude \\
POSEPOCH & Position epoch \\
F0 & Frequency of pulsar rotation \\
F1 & Derivative of frequency \\
PEPOCH & Period epoch \\
DM & Dispersion measure \\
EPHEM &  Which solar system ephemeris to use \\
CLK &  Definition of clock to use \\
TZRSITE & Telescope site code corresponding to TZR (first site TOA after PEPOCH) \\
F2 & Second frequency derivative \\
\end{tabular}
\caption{Pulsar parameters kept in the stripped par files which were used to create simulated TOAs with \texttt{Tempo2}.}
\label{tab:kept_params}
\end{table}

\begin{table}[ht]
\centering
\begin{tabular}{lll}
\hline
Parameter & Details & Value \\
\hline
NDOBS & Number of days between observations & 28 \\
NOBSD & Number of observations per day & 1 \\
RANDHA & Random hour angle (y/n) & n \\
START & Start date of observation (MJD) & Varied \\
END & End date of observation (MJD) & 58850 (Jan 2020) \\
RMS & RMS noise to be added (ms) & Varied \\
\end{tabular}
\caption{Run parameters used to generate residuals. Unless otherwise stated, these parameters were kept constant between pulsars.}
\label{tab:run_params}
\end{table}

\begin{table}[ht]
    \centering
    \begin{tabular}{c c c c c c c c c c}
\hline
Name & F0 & F0 err & F1 & F1 err & F2 & RMS & Mean n$^*$ & Std n$^*$ \\
 & (s) & (s) & & (s$^{-1}$) & (s$^{-1}$) & ms & & \\ 
\hline
J0023+0923 & 327.8 & 3.45e-13 & -1.23e-15 & 1.49e-20 & 2.30e-32 & 0.000288 & 5.17 & 2.83 \\
J0030+0451 & 205.5 & 7.74e-11 & -4.30e-16 & 2.78e-19 & 4.49e-33 & 0.0002 & 5.03 & 2.83 \\
J0340+4130 & 303.1 & 9.67e-13 & -6.47e-16 & 4.17e-20 & 6.91e-33 & 0.000449 & 4.79 & 2.82 \\
J0613-0200 & 326.6 & 1.63e-12 & -1.02e-15 & 1.71e-20 & 1.60e-32 & 0.000188 & 4.67 & 2.80 \\
J0636+5128 & 348.6 & 8.68e-13 & -4.19e-16 & 7.25e-20 & 2.52e-33 & 0.000596 & 5.05 & 2.83 \\
J0645+5158 & 112.9 & 1.23e-13 & -6.28e-17 & 4.49e-21 & 1.74e-34 & 0.000196 & 5.10 & 2.82 \\
J0740+6620 & 346.5 & 1.18e-12 & -1.46e-15 & 7.91e-20 & 3.09e-32 & 0.000106 & 4.49 & 2.80 \\
J0931-1902 & 215.6 & 8.77e-13 & -1.69e-16 & 5.35e-20 & 6.60e-34 & 0.000424 & 5.08 & 2.84 \\
J1012+5307 & 190.3 & 1.49e-12 & -6.20e-16 & 1.75e-20 & 1.01e-32 & 0.000209 & 5.17 & 2.82 \\
J1024-0719 & 193.7 & 6.18e-13 & -6.96e-16 & 8.38e-21 & 1.25e-32 & 0.00024 & 5.19 & 2.83 \\
J1125+7819 & 238.0 & 2.35e-12 & -3.93e-16 & 1.61e-19 & 3.25e-33 & 0.000614 & 5.02 & 2.88 \\
J1453+1902 & 172.6 & 2.09e-12 & -3.47e-16 & 1.15e-19 & 3.50e-33 & 0.000798 & 5.02 & 2.85 \\
J1455-3330 & 125.2 & 3.59e-13 & -3.81e-16 & 4.40e-21 & 5.80e-33 & 0.000544 & 4.94 & 2.85 \\
J1600-3053 & 277.9 & 1.58e-13 & -7.34e-16 & 3.63e-21 & 9.69e-33 & 0.000213 & 4.99 & 2.80 \\
J1614-2230 & 317.4 & 2.21e-13 & -9.69e-16 & 5.20e-21 & 1.48e-32 & 0.000175 & 5.06 & 2.85 \\
J1640+2224 & 316.1 & 1.50e-13 & -2.82e-16 & 2.53e-21 & 1.25e-33 & 0.000142 & 5.13 & 2.79 \\
J1643-1224 & 216.4 & 4.21e-12 & -8.64e-16 & 5.39e-20 & 1.73e-32 & 0.000292 & 5.34 & 2.80 \\
J1713+0747 & 218.8 & 7.93e-14 & -4.08e-16 & 1.11e-21 & 3.81e-33 & 8.1e-05 & 4.75 & 2.86 \\
J1738+0333 & 170.9 & 4.12e-13 & -7.05e-16 & 1.08e-20 & 1.45e-32 & 0.000272 & 4.54 & 2.83 \\
J1741+1351 & 266.9 & 2.13e-13 & -2.15e-15 & 8.23e-21 & 8.68e-32 & 0.000148 & 6.77 & 2.36 \\
J1744-1134 & 245.4 & 1.78e-12 & -5.38e-16 & 1.68e-20 & 5.90e-33 & 0.000278 & 4.99 & 2.83 \\
J1747-4036 & 607.7 & 2.03e-10 & -4.85e-15 & 2.21e-18 & 1.94e-31 & 0.000767 & 5.25 & 2.81 \\
J1832-0836 & 367.8 & 5.82e-13 & -1.12e-15 & 3.16e-20 & 1.70e-32 & 0.000195 & 5.10 & 2.82 \\
J1853+1303 & 244.4 & 2.79e-12 & -5.21e-16 & 5.36e-20 & 5.55e-33 & 9.2e-05 & 5.15 & 2.88 \\
B1855+09 & 186.5 & 4.05e-12 & -6.20e-16 & 2.41e-20 & 1.03e-32 & 0.000357 & 5.13 & 2.84 \\
J1903+0327 & 465.1 & 2.03e-11 & -4.07e-15 & 3.70e-19 & 1.78e-31 & 0.000394 & 4.93 & 2.82 \\
J1909-3744 & 339.3 & 1.21e-12 & -1.61e-15 & 9.76e-21 & 3.84e-32 & 5.8e-05 & 3.48 & 2.50 \\
J1910+1256 & 200.7 & 3.83e-13 & -3.90e-16 & 8.10e-21 & 3.79e-33 & 0.000399 & 5.03 & 2.83 \\
J1911+1347 & 216.2 & 2.15e-13 & -7.91e-16 & 1.31e-20 & 1.45e-32 & 0.000115 & 5.28 & 2.77 \\
J1918-0642 & 130.8 & 1.11e-13 & -4.39e-16 & 1.45e-21 & 7.38e-33 & 0.000296 & 4.83 & 2.85 \\
J1923+2515 & 264.0 & 4.93e-13 & -6.66e-16 & 1.41e-20 & 8.40e-33 & 0.000237 & 4.95 & 2.83 \\
B1937+21 & 641.9 & 3.26e-11 & -4.33e-14 & 1.79e-19 & 1.46e-29 & 0.000103 & 5.01 & 0.06 \\
J1944+0907 & 192.9 & 4.13e-13 & -6.45e-16 & 6.20e-21 & 1.08e-32 & 0.000375 & 5.11 & 2.80 \\
J1946+3417 & 315.4 & 6.12e-12 & -3.15e-16 & 2.18e-19 & 1.57e-33 & 0.000143 & 4.90 & 2.83 \\
B1953+29 & 163.0 & 7.93e-13 & -7.91e-16 & 2.76e-20 & 1.92e-32 & 0.000475 & 5.33 & 2.85 \\
J2010-1323 & 191.5 & 1.48e-13 & -1.77e-16 & 4.31e-21 & 8.17e-34 & 0.00025 & 4.95 & 2.82 \\
J2017+0603 & 345.3 & 5.14e-13 & -9.53e-16 & 3.30e-20 & 1.32e-32 & 9.7e-05 & 4.42 & 2.81 \\
J2033+1734 & 168.1 & 9.20e-13 & -3.15e-16 & 5.71e-20 & 2.95e-33 & 0.00052 & 4.93 & 2.87 \\
J2043+1711 & 420.2 & 3.04e-13 & -9.26e-16 & 9.33e-21 & 1.02e-32 & 0.000122 & 4.69 & 2.78 \\
J2145-0750 & 62.3  & 9.24e-13 & -1.16e-16 & 8.99e-21 & 1.07e-33 & 0.000274 & 4.95 & 2.82 \\
J2214+3000 & 320.6 & 1.13e-12 & -1.51e-15 & 3.71e-20 & 3.57e-32 & 0.000419 & 5.17 & 2.87 \\
J2229+2643 & 335.8 & 7.18e-13 & -1.72e-16 & 4.84e-20 & 4.40e-34 & 0.000196 & 5.01 & 2.87 \\
J2234+0611 & 279.6 & 3.94e-13 & -9.39e-16 & 2.25e-20 & 1.58e-32 & 3.5e-05 & 4.21 & 2.68 \\
J2234+0944 & 275.7 & 8.77e-13 & -1.53e-15 & 3.77e-20 & 4.23e-32 & 0.000165 & 5.22 & 2.80 \\
J2302+4442 & 192.6 & 7.64e-13 & -5.14e-16 & 3.52e-20 & 6.87e-33 & 0.000693 & 4.95 & 2.84 \\
J2317+1439 & 290.3 & 3.20e-11 & -2.05e-16 & 1.14e-19 & 7.22e-34 & 0.000204 & 4.97 & 2.85 \\
J2322+2057 & 208.0 & 9.85e-13 & -4.18e-16 & 1.02e-19 & 4.19e-33 & 0.000237 & 5.11 & 2.85 \\
\hline
    \end{tabular}
    \caption{Pulsar values used in this analysis. Columns denoted by $^*$ include values calculated from a simulated 20 year run with an average RMS of \scinum{1}{-5} ms scaled using the listed RMS value and equation \eqref{eq:rms} and assuming all pulsars have a braking index of exactly $n=5$.}
    \label{tab:pul_vals}
\end{table}

\pagebreak

\section{TOAs and posteriors}
\label{sec:posteriors}

\begin{figure*}[ht]
    \centering
    \includegraphics[width=\textwidth]{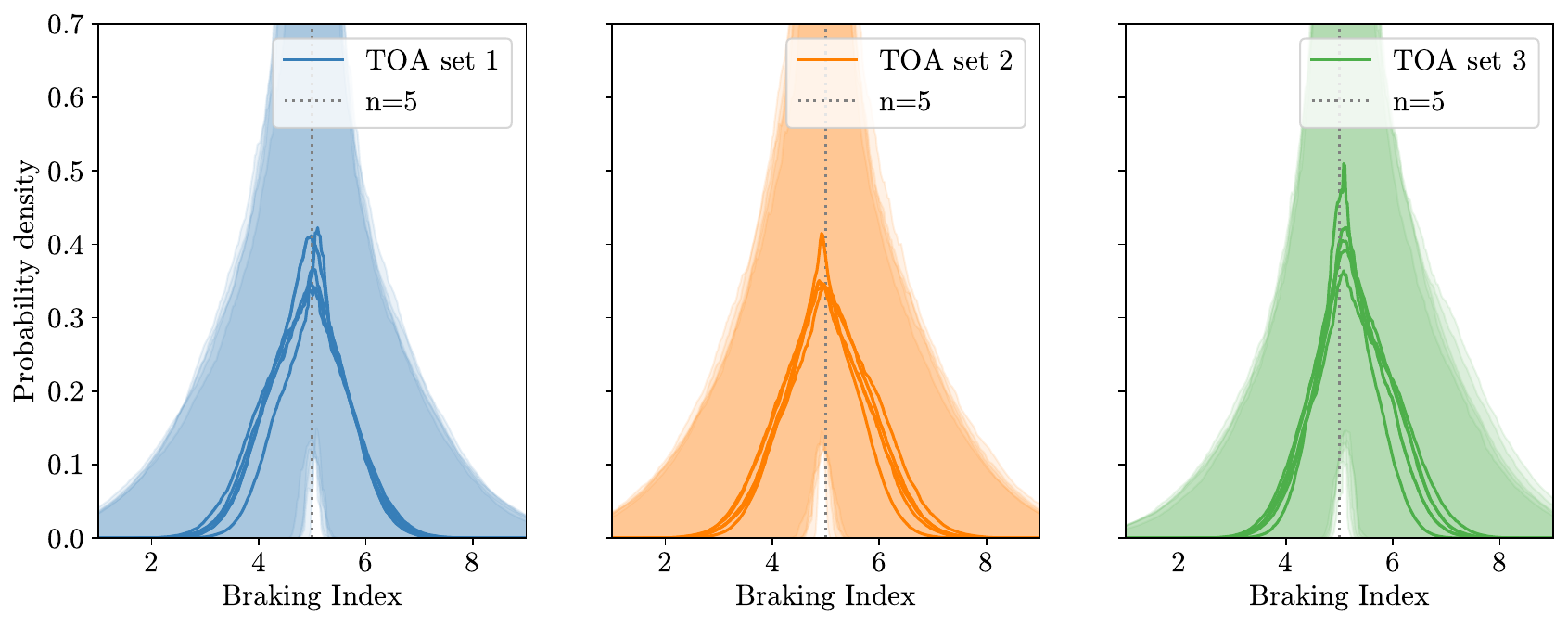}
    \caption{The result of identical analyses performed on three sets of TOAs produced by \texttt{Tempo2} given identical parameters. It demonstrates the variation introduced during the TOA generation step and posterior analysis. The solid lines represent the median values while the shaded regions are bounded by the 5th and 95th percentiles. The mean $\text{OR}_{4-6}$ for the blue, orange, and green TOAs are: $0.96\pm0.03$, $0.94\pm0.02$, and $1.04\pm0.06$ respectively.}
    \label{fig:TOAtests}
\end{figure*}

In \figref{fig:TOAtests}, we look into the variation caused at different stages in the analysis. Each colour represents a different set of TOAs produced by \texttt{Tempo2} given identical parameters. These three sets of TOAs are then put through the rest of the analysis five times to see the variation introduced when obtaining posteriors. The variation on each TOA itself is produced during either the \texttt{enterprise} or \texttt{posteriorstacker} stage while the differences between each TOA would be introduced by \texttt{Tempo2}. We calculate the mean and standard deviation of $\text{OR}_{4-6}$ for each TOA set to be $0.96\pm0.03$, $0.94\pm0.02$, and $1.04\pm0.06$ for blue, orange, and green respectively. While the green TOA set had higher odds ratios overall, its lowest was still well within the range of the other two. Blue and orange are additionally within each other's ranges, suggesting that the variation from postprocessing dominates the variation from the TOA set used.

We conclude that there is little correlation between the TOA data used and the probability density after the full analysis pipeline. Each set of TOAs demonstrates a similar amount of variation with no TOA producing consistently better results. To rule out the amount of variation introduced by \texttt{posteriorstacker}, the script is additionally run repeatedly on the same posteriors. This results in negligible variation, so it is concluded that the majority of the variation is introduced when using \texttt{enterprise} to obtain posteriors on $n$.

\pagebreak

\section{Varying braking index}
\label{sec:vary-n}

\begin{figure}[ht]
    \centering
    \includegraphics[width=0.473\textwidth]{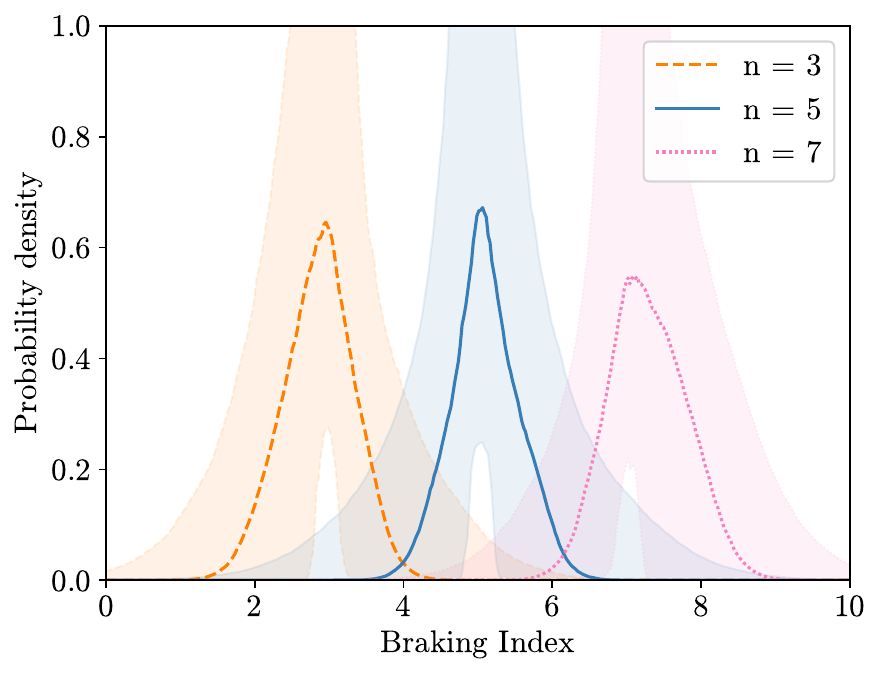}
    \caption{Analyses with identical parameters aside from varying the braking index used to set $\Ddot{f}$ when generating the simulate TOAs. The solid lines represent the median values while the shaded regions are bounded by the 5th and 95th percentiles.}
    \label{fig:vary-n}
\end{figure}

So far all analyses have been run on simulated TOAs with $n$ fixed at 5. In order to test this method's ability to identify any value of $n$, runs are performed on TOAs with different $n$. We perform  three analyses with 20 years of observation at an RMS noise level of \scinum{1}{-6}. These values are chosen as they produce accurate and confident results in the other analyses. Tests for $n=3$ and $n=7$ are chosen as they represent the braking indices due to magnetic dipole radiation and r-mode GWs respectively (see \secref{sec:intro}). \figref{fig:vary-n} shows that a strong signal was found for all three analyses with similar confidences and accuracies, verifying that this method is not biased towards a braking index of 5. 

\bibliography{pulsar-braking}{}

\begin{thebibliography}{}
\expandafter\ifx\csname natexlab\endcsname\relax\def\natexlab#1{#1}\fi
\providecommand{\url}[1]{\href{#1}{#1}}
\providecommand{\dodoi}[1]{doi:~\href{http://doi.org/#1}{\nolinkurl{#1}}}
\providecommand{\doeprint}[1]{\href{http://ascl.net/#1}{\nolinkurl{http://ascl.net/#1}}}
\providecommand{\doarXiv}[1]{\href{https://arxiv.org/abs/#1}{\nolinkurl{https://arxiv.org/abs/#1}}}

\bibitem[{{Abbott} {et~al.}(2021{\natexlab{a}}){Abbott}, {Abbott}, {Abraham}, {Acernese}, {Ackley}, {Adams}, {Adams}, {Adhikari}, {Adya}, {Affeldt}, {Agathos}, {Agatsuma}, {Aggarwal}, {Aguiar}, {Aiello}, {Ain}, {Ajith}, {Akcay}, {Allen}, {Allocca}, {Altin}, {Amato}, {Anand}, {Ananyeva}, {Anderson}, {Anderson}, {Angelova}, {Ansoldi}, {Antelis}, {Antier}, {Appert}, {Arai}, {et~al.}}]{abbott2021gwtc2}
{Abbott}, R., {Abbott}, T.~D., {Abraham}, S., {et~al.} 2021{\natexlab{a}}, PhRvX, 11, 021053, \dodoi{10.1103/PhysRevX.11.021053}

\bibitem[{{Abbott} {et~al.}(2021{\natexlab{b}}){Abbott}, {Abbott}, {Abraham}, {Acernese}, {Ackley}, {Adams}, {Adams}, {Adhikari}, {Adya}, {Affeldt}, \& et~al.}]{2021ApJ...915L...5A}
---. 2021{\natexlab{b}}, \apjl, 915, L5, \dodoi{10.3847/2041-8213/ac082e}

\bibitem[{{Abbott} {et~al.}(2021{\natexlab{c}}){Abbott}, {Abbott}, {Abraham}, {Acernese}, {Ackley}, {Adams}, {Adams}, {Adhikari}, {Adya}, {Affeldt}, \& et~al.}]{2021ApJ...922...71A}
---. 2021{\natexlab{c}}, \apj, 922, 71, \dodoi{10.3847/1538-4357/ac0d52}

\bibitem[{{Abbott} {et~al.}(2023){Abbott}, {Abbott}, {Acernese}, {Ackley}, {Adams}, {Adhikari}, {Adhikari}, {Adya}, {Affeldt}, {Agarwal}, {Agathos}, {Agatsuma}, {Aggarwal}, {Aguiar}, {Aiello}, \& et~al.}]{2023PhRvX..13d1039A}
{Abbott}, R., {Abbott}, T.~D., {Acernese}, F., {et~al.} 2023, Physical Review X, 13, 041039, \dodoi{10.1103/PhysRevX.13.041039}

\bibitem[{{Agazie} {et~al.}(2023){Agazie}, {Anumarlapudi}, {Archibald}, {Arzoumanian}, {Baker}, {B{\'e}csy}, {Blecha}, {Brazier}, {Brook}, {Burke-Spolaor}, {Burnette}, {Case}, {Charisi}, {Chatterjee}, {Chatziioannou}, {Cheeseboro}, {Chen}, {Cohen}, {Cordes}, {Cornish}, {Crawford}, {Cromartie}, {Crowter}, {Cutler}, {Decesar}, {Degan}, {Demorest}, {Deng}, {Dolch}, {Drachler}, {Ellis}, {Ferrara}, {Fiore}, {Fonseca}, {Freedman}, {Garver-Daniels}, {Gentile}, {Gersbach}, {Glaser}, {Good}, {G{\"u}ltekin}, {Hazboun}, {Hourihane}, {Islo}, {Jennings}, {Johnson}, {Jones}, {Kaiser}, {Kaplan}, {Kelley}, {Kerr}, {Key}, {Klein}, {Laal}, {Lam}, {Lamb}, {Lazio}, {Lewandowska}, {Littenberg}, {Liu}, {Lommen}, {Lorimer}, {Luo}, {Lynch}, {Ma}, {Madison}, {Mattson}, {McEwen}, {McKee}, {McLaughlin}, {McMann}, {Meyers}, {Meyers}, {Mingarelli}, {Mitridate}, {Natarajan}, {Ng}, {Nice}, {Ocker}, {Olum}, {Pennucci}, {Perera}, {Petrov}, {Pol}, {Radovan}, {Ransom}, {Ray}, {Romano}, {Sardesai}, {Schmiedekamp}, {Schmiedekamp}, {Schmitz},
  {Schult}, {Shapiro-Albert}, {Siemens}, {Simon}, {Siwek}, {Stairs}, {Stinebring}, {Stovall}, {Sun}, {Susobhanan}, {Swiggum}, {Taylor}, {Taylor}, {Turner}, {Unal}, {Vallisneri}, {van Haasteren}, {Vigeland}, {Wahl}, {Wang}, {Witt}, {Young}, \& {Nanograv Collaboration}}]{2023ApJ...951L...8A}
{Agazie}, G., {Anumarlapudi}, A., {Archibald}, A.~M., {et~al.} 2023, \apjl, 951, L8, \dodoi{10.3847/2041-8213/acdac6}

\bibitem[{{Alam} {et~al.}(2021{\natexlab{a}}){Alam}, {Arzoumanian}, {Baker}, {Blumer}, {Bohler}, {Brazier}, {Brook}, {Burke-Spolaor}, {Caballero}, {Camuccio}, {Chamberlain}, {Chatterjee}, {Cordes}, {Cornish}, {Crawford}, {Cromartie}, {Decesar}, {Demorest}, {Dolch}, {Ellis}, {Ferdman}, {Ferrara}, {Fiore}, {Fonseca}, {Garcia}, {Garver-Daniels}, {Gentile}, {Good}, {Gusdorff}, {Halmrast}, {Hazboun}, {Islo}, {Jennings}, {Jessup}, {Jones}, {Kaiser}, {Kaplan}, {Kelley}, {Key}, {Lam}, {Lazio}, {Lorimer}, {Luo}, {Lynch}, {Madison}, {Maraccini}, {McLaughlin}, {Mingarelli}, {Ng}, {Nguyen}, {Nice}, {Pennucci}, {Pol}, {Ramette}, {Ransom}, {Ray}, {Shapiro-Albert}, {Siemens}, {Simon}, {Spiewak}, {Stairs}, {Stinebring}, {Stovall}, {Swiggum}, {Taylor}, {Tripepi}, {Vallisneri}, {Vigeland}, {Witt}, {Zhu}, \& {Nanograv Collaboration}}]{nanograv_wide}
{Alam}, M.~F., {Arzoumanian}, Z., {Baker}, P.~T., {et~al.} 2021{\natexlab{a}}, \apjs, 252, 5, \dodoi{10.3847/1538-4365/abc6a1}

\bibitem[{{Alam} {et~al.}(2021{\natexlab{b}}){Alam}, {Arzoumanian}, {Baker}, {Blumer}, {Bohler}, {Brazier}, {Brook}, {Burke-Spolaor}, {Caballero}, {Camuccio}, {Chamberlain}, {Chatterjee}, {Cordes}, {Cornish}, {Crawford}, {Cromartie}, {Decesar}, {Demorest}, {Dolch}, {Ellis}, {Ferdman}, {Ferrara}, {Fiore}, {Fonseca}, {Garcia}, {Garver-Daniels}, {Gentile}, {Good}, {Gusdorff}, {Halmrast}, {Hazboun}, {Islo}, {Jennings}, {Jessup}, {Jones}, {Kaiser}, {Kaplan}, {Kelley}, {Key}, {Lam}, {Lazio}, {Lorimer}, {Luo}, {Lynch}, {Madison}, {Maraccini}, {McLaughlin}, {Mingarelli}, {Ng}, {Nguyen}, {Nice}, {Pennucci}, {Pol}, {Ramette}, {Ransom}, {Ray}, {Shapiro-Albert}, {Siemens}, {Simon}, {Spiewak}, {Stairs}, {Stinebring}, {Stovall}, {Swiggum}, {Taylor}, {Tripepi}, {Vallisneri}, {Vigeland}, {Witt}, {Zhu}, \& {Nanograv Collaboration}}]{nanograv_narrow}
---. 2021{\natexlab{b}}, \apjs, 252, 4, \dodoi{10.3847/1538-4365/abc6a0}

\bibitem[{{Andersson}(1998)}]{1998ApJ...502..708A}
{Andersson}, N. 1998, \apj, 502, 708, \dodoi{10.1086/305919}

\bibitem[{{Ashton} \& {Talbot}(2021)}]{bilby_mcmc}
{Ashton}, G., \& {Talbot}, C. 2021, \mnras, 507, 2037, \dodoi{10.1093/mnras/stab2236}

\bibitem[{{Ashton} {et~al.}(2019){Ashton}, {H{\"u}bner}, {Lasky}, {Talbot}, {Ackley}, {Biscoveanu}, {Chu}, {Divakarla}, {Easter}, {Goncharov}, {Hernandez Vivanco}, {Harms}, {Lower}, {Meadors}, {Melchor}, {Payne}, {Pitkin}, {Powell}, {Sarin}, {Smith}, \& {Thrane}}]{bilby1}
{Ashton}, G., {H{\"u}bner}, M., {Lasky}, P.~D., {et~al.} 2019, \apjs, 241, 27, \dodoi{10.3847/1538-4365/ab06fc}

\bibitem[{{Baronchelli} {et~al.}(2020){Baronchelli}, {Nandra}, \& {Buchner}}]{posterior_method}
{Baronchelli}, L., {Nandra}, K., \& {Buchner}, J. 2020, \mnras, 498, 5284, \dodoi{10.1093/mnras/staa2684}

\bibitem[{{Bonazzola} \& {Gourgoulhon}(1996)}]{1996AA...312..675B}
{Bonazzola}, S., \& {Gourgoulhon}, E. 1996, \aap, 312, 675

\bibitem[{{Buchner}(2021)}]{posteriorstacker}
{Buchner}, J. 2021.
\newblock \url{https://github.com/JohannesBuchner/PosteriorStacker}

\bibitem[{{Chen} {et~al.}(2021){Chen}, {Caballero}, {Guo}, {Chalumeau}, {Liu}, {Shaifullah}, {Lee}, {Babak}, {Desvignes}, {Parthasarathy}, {Hu}, {van der Wateren}, {Antoniadis}, {Bak Nielsen}, {Bassa}, {Berthereau}, {Burgay}, {Champion}, {Cognard}, {Falxa}, {Ferdman}, {Freire}, {Gair}, {Graikou}, {Guillemot}, {Jang}, {Janssen}, {Karuppusamy}, {Keith}, {Kramer}, {Liu}, {Lyne}, {Main}, {McKee}, {Mickaliger}, {Perera}, {Perrodin}, {Petiteau}, {Porayko}, {Possenti}, {Samajdar}, {Sanidas}, {Sesana}, {Speri}, {Stappers}, {Theureau}, {Tiburzi}, {Vecchio}, {Verbiest}, {Wang}, {Wang}, \& {Xu}}]{24years}
{Chen}, S., {Caballero}, R.~N., {Guo}, Y.~J., {et~al.} 2021, \mnras, 508, 4970, \dodoi{10.1093/mnras/stab2833}

\bibitem[{{Cutler}(2002)}]{2002PhRvD..66h4025C}
{Cutler}, C. 2002, \prd, 66, 084025, \dodoi{10.1103/PhysRevD.66.084025}

\bibitem[{{Edwards} {et~al.}(2006){Edwards}, {Hobbs}, \& {Manchester}}]{tempo22}
{Edwards}, R.~T., {Hobbs}, G.~B., \& {Manchester}, R.~N. 2006, \mnras, 372, 1549, \dodoi{10.1111/j.1365-2966.2006.10870.x}

\bibitem[{Ellis {et~al.}(2020)Ellis, Vallisneri, Taylor, \& Baker}]{enterprise}
Ellis, J.~A., Vallisneri, M., Taylor, S.~R., \& Baker, P.~T. 2020, ENTERPRISE: Enhanced Numerical Toolbox Enabling a Robust PulsaR Inference SuitE, Zenodo, \dodoi{10.5281/zenodo.4059815}

\bibitem[{{EPTA Collaboration and InPTA Collaboration} {et~al.}(2023){EPTA Collaboration and InPTA Collaboration}, {Antoniadis, J.}, {Arumugam, P.}, {Arumugam, S.}, {Babak, S.}, {Bagchi, M.}, {Bak Nielsen, A.-S.}, {Bassa, C. G.}, {Bathula, A.}, {Berthereau, A.}, {Bonetti, M.}, {Bortolas, E.}, {Brook, P. R.}, {Burgay, M.}, {Caballero, R. N.}, {Chalumeau, A.}, {Champion, D. J.}, {Chanlaridis, S.}, {Chen, S.}, {Cognard, I.}, {Dandapat, S.}, {Deb, D.}, {Desai, S.}, {Desvignes, G.}, {Dhanda-Batra, N.}, {Dwivedi, C.}, {Falxa, M.}, {Ferdman, R. D.}, {Franchini, A.}, {Gair, J. R.}, {Goncharov, B.}, {Gopakumar, A.}, {Graikou, E.}, {Grießmeier, J.-M.}, {Guillemot, L.}, {Guo, Y. J.}, {Gupta, Y.}, {Hisano, S.}, {Hu, H.}, {Iraci, F.}, {Izquierdo-Villalba, D.}, {Jang, J.}, {Jawor, J.}, {Janssen, G. H.}, {Jessner, A.}, {Joshi, B. C.}, {Kareem, F.}, {Karuppusamy, R.}, {Keane, E. F.}, {Keith, M. J.}, {Kharbanda, D.}, {Kikunaga, T.}, {Kolhe, N.}, {Kramer, M.}, {Krishnakumar, M. A.}, {Lackeos, K.}, {Lee, K. J.}, {Liu, K.},
  {Liu, Y.}, {Lyne, A. G.}, {McKee, J. W.}, {Maan, Y.}, {Main, R. A.}, {Mickaliger, M. B.}, {Niţu, I. C.}, {Nobleson, K.}, {Paladi, A. K.}, {Parthasarathy, A.}, {Perera, B. B. P.}, {Perrodin, D.}, {Petiteau, A.}, {Porayko, N. K.}, {Possenti, A.}, {Prabu, T.}, {Quelquejay Leclere, H.}, {Rana, P.}, {Samajdar, A.}, {Sanidas, S. A.}, {Sesana, A.}, {Shaifullah, G.}, {Singha, J.}, {Speri, L.}, {Spiewak, R.}, {Srivastava, A.}, {Stappers, B. W.}, {Surnis, M.}, {Susarla, S. C.}, {Susobhanan, A.}, {Takahashi, K.}, {Tarafdar, P.}, {Theureau, G.}, {Tiburzi, C.}, {van der Wateren, E.}, {Vecchio, A.}, {Venkatraman Krishnan, V.}, {Verbiest, J. P. W.}, {Wang, J.}, {Wang, L.}, \& {Wu, Z.}}]{refId0}
{EPTA Collaboration and InPTA Collaboration}, {Antoniadis, J.}, {Arumugam, P.}, {et~al.} 2023, A\&A, 678, A50, \dodoi{10.1051/0004-6361/202346844}

\bibitem[{{Friedman} \& {Morsink}(1998)}]{1998ApJ...502..714F}
{Friedman}, J.~L., \& {Morsink}, S.~M. 1998, \apj, 502, 714, \dodoi{10.1086/305920}

\bibitem[{{Gittins} \& {Andersson}(2021)}]{2021MNRAS.507..116G}
{Gittins}, F., \& {Andersson}, N. 2021, \mnras, 507, 116, \dodoi{10.1093/mnras/stab2048}

\bibitem[{{Goncharov}(2021)}]{entwarp}
{Goncharov}, B. 2021.
\newblock \url{https://github.com/bvgoncharov/enterprise_warp}

\bibitem[{Goncharov {et~al.}(2024)Goncharov, Zic, Reardon, \& Pitkin}]{enterprise_warp_edit}
Goncharov, B., Zic, A., Reardon, D., \& Pitkin, M. 2024, mattpitkin/enterprise\_warp: Braking index sampling, braking\_index\_sampling,  Zenodo, \dodoi{10.5281/zenodo.13274448}

\bibitem[{Hamil {et~al.}(2015)Hamil, Stone, Urbanec, \& Urbancová}]{Hamil_2015}
Hamil, O., Stone, J., Urbanec, M., \& Urbancová, G. 2015, Physical Review D, 91, \dodoi{10.1103/physrevd.91.063007}

\bibitem[{{Harris} {et~al.}(2020){Harris}, {Millman}, {van der Walt}, {Gommers}, {Virtanen}, {Cournapeau}, {Wieser}, {Taylor}, {Berg}, {Smith}, {Kern}, {Picus}, {Hoyer}, {van Kerkwijk}, {Brett}, {Haldane}, {del R{\'\i}o}, {Wiebe}, {Peterson}, {G{\'e}rard-Marchant}, {Sheppard}, {Reddy}, {Weckesser}, {Abbasi}, {Gohlke}, \& {Oliphant}}]{numpy}
{Harris}, C.~R., {Millman}, K.~J., {van der Walt}, S.~J., {et~al.} 2020, \nat, 585, 357, \dodoi{10.1038/s41586-020-2649-2}

\bibitem[{{Hobbs} {et~al.}(2006){Hobbs}, {Edwards}, \& {Manchester}}]{tempo21}
{Hobbs}, G., {Edwards}, R., \& {Manchester}, R. 2006, Chinese Journal of Astronomy and Astrophysics Supplement, 6, 189

\bibitem[{{Hobbs} {et~al.}(2009){Hobbs}, {Jenet}, {Lee}, {Verbiest}, {Yardley}, {Manchester}, {Lommen}, {Coles}, {Edwards}, \& {Shettigara}}]{tempo23}
{Hobbs}, G., {Jenet}, F., {Lee}, K.~J., {et~al.} 2009, \mnras, 394, 1945, \dodoi{10.1111/j.1365-2966.2009.14391.x}

\bibitem[{{Hunter}(2007)}]{2007CSE.....9...90H}
{Hunter}, J.~D. 2007, CSE, 9, 90, \dodoi{10.1109/MCSE.2007.55}

\bibitem[{{Lam}(2018)}]{cadence}
{Lam}, M.~T. 2018, \apj, 868, 33, \dodoi{10.3847/1538-4357/aae533}

\bibitem[{{Liu} {et~al.}(2011){Liu}, {Verbiest}, {Kramer}, {Stappers}, {van Straten}, \& {Cordes}}]{2011MNRAS.417.2916L}
{Liu}, K., {Verbiest}, J.~P.~W., {Kramer}, M., {et~al.} 2011, \mnras, 417, 2916, \dodoi{10.1111/j.1365-2966.2011.19452.x}

\bibitem[{{Liu} {et~al.}(2019){Liu}, {Keith}, {Bassa}, \& {Stappers}}]{Liu19}
{Liu}, X.~J., {Keith}, M.~J., {Bassa}, C.~G., \& {Stappers}, B.~W. 2019, \mnras, 488, 2190, \dodoi{10.1093/mnras/stz1801}

\bibitem[{{Lower} {et~al.}(2021){Lower}, {Johnston}, {Dunn}, {Shannon}, {Bailes}, {Dai}, {Kerr}, {Manchester}, {Melatos}, {Oswald}, {Parthasarathy}, {Sobey}, \& {Weltevrede}}]{glitches}
{Lower}, M.~E., {Johnston}, S., {Dunn}, L., {et~al.} 2021, \mnras, 508, 3251, \dodoi{10.1093/mnras/stab2678}

\bibitem[{{Ostriker} \& {Gunn}(1969)}]{1969ApJ...157.1395O}
{Ostriker}, J.~P., \& {Gunn}, J.~E. 1969, \apj, 157, 1395, \dodoi{10.1086/150160}

\bibitem[{{Palomba}(2000)}]{2000A&A...354..163P}
{Palomba}, C. 2000, \aap, 354, 163

\bibitem[{{Palomba}(2005)}]{2005palomba}
---. 2005, \mnras, 359, 1150, \dodoi{10.1111/j.1365-2966.2005.08975.x}

\bibitem[{{Parthasarathy} {et~al.}(2020){Parthasarathy}, {Johnston}, {Shannon}, {Lentati}, {Bailes}, {Dai}, {Kerr}, {Manchester}, {Os{\l}owski}, {Sobey}, {van Straten}, \& {Weltevrede}}]{obs_n}
{Parthasarathy}, A., {Johnston}, S., {Shannon}, R.~M., {et~al.} 2020, \mnras, 494, 2012, \dodoi{10.1093/mnras/staa882}

\bibitem[{{Parthasarathy} {et~al.}(2021){Parthasarathy}, {Bailes}, {Shannon}, {van Straten}, {Os{\l}owski}, {Johnston}, {Spiewak}, {Reardon}, {Kramer}, {Venkatraman Krishnan}, {Pennucci}, {Abbate}, {Buchner}, {Camilo}, {Champion}, {Geyer}, {Hugo}, {Jameson}, {Karastergiou}, {Keith}, \& {Serylak}}]{2021MNRAS.502..407P}
{Parthasarathy}, A., {Bailes}, M., {Shannon}, R.~M., {et~al.} 2021, \mnras, 502, 407, \dodoi{10.1093/mnras/stab037}

\bibitem[{{Perrodin} {et~al.}(2013){Perrodin}, {Jenet}, {Lommen}, {Finn}, {Demorest}, {Ferdman}, {Gonzalez}, {Nice}, {Ransom}, \& {Stairs}}]{noise2}
{Perrodin}, D., {Jenet}, F., {Lommen}, A., {et~al.} 2013, arXiv e-prints, arXiv:1311.3693.
\newblock \doarXiv{1311.3693}

\bibitem[{{Price-Whelan} {et~al.}(2018){Price-Whelan}, {Sip{\H{o}}cz}, {G{\"u}nther}, {et~al.}}]{astropy:2018}
{Price-Whelan}, A.~M., {Sip{\H{o}}cz}, B.~M., {G{\"u}nther}, H.~M., {et~al.} 2018, \aj, 156, 123, \dodoi{10.3847/1538-3881/aabc4f}

\bibitem[{{Reardon} {et~al.}(2023){Reardon}, {Zic}, {Shannon}, {Hobbs}, {Bailes}, {Di Marco}, {Kapur}, {Rogers}, {Thrane}, {Askew}, {Bhat}, {Cameron}, {Cury{\l}o}, {Coles}, {Dai}, {Goncharov}, {Kerr}, {Kulkarni}, {Levin}, {Lower}, {Manchester}, {Mandow}, {Miles}, {Nathan}, {Os{\l}owski}, {Russell}, {Spiewak}, {Zhang}, \& {Zhu}}]{2023ApJ...951L...6R}
{Reardon}, D.~J., {Zic}, A., {Shannon}, R.~M., {et~al.} 2023, \apjl, 951, L6, \dodoi{10.3847/2041-8213/acdd02}

\bibitem[{{Robitaille} {et~al.}(2013){Robitaille}, {Tollerud}, {Greenfield}, {et~al.}}]{astropy:2013}
{Robitaille}, T.~P., {Tollerud}, E.~J., {Greenfield}, P., {et~al.} 2013, \aap, 558, A33, \dodoi{10.1051/0004-6361/201322068}

\bibitem[{{Romero-Shaw} {et~al.}(2020){Romero-Shaw}, {Talbot}, {Biscoveanu}, {D'Emilio}, {Ashton}, {Berry}, {Coughlin}, {Galaudage}, {Hoy}, {H{\"u}bner}, {Phukon}, {Pitkin}, {Rizzo}, {Sarin}, {Smith}, {Stevenson}, {Vajpeyi}, {Ar{\`e}ne}, {Athar}, {Banagiri}, {Bose}, {Carney}, {Chatziioannou}, {Clark}, {Colleoni}, {Cotesta}, {Edelman}, {Estell{\'e}s}, {Garc{\'\i}a-Quir{\'o}s}, {Ghosh}, {Green}, {Haster}, {Husa}, {Keitel}, {Kim}, {Hernandez-Vivanco}, {Maga{\~n}a Hernandez}, {Karathanasis}, {Lasky}, {De Lillo}, {Lower}, {Macleod}, {Mateu-Lucena}, {Miller}, {Millhouse}, {Morisaki}, {Oh}, {Ossokine}, {Payne}, {Powell}, {Pratten}, {P{\"u}rrer}, {Ramos-Buades}, {Raymond}, {Thrane}, {Veitch}, {Williams}, {Williams}, \& {Xiao}}]{bilby2}
{Romero-Shaw}, I.~M., {Talbot}, C., {Biscoveanu}, S., {et~al.} 2020, \mnras, 499, 3295, \dodoi{10.1093/mnras/staa2850}

\bibitem[{{Speagle}(2020)}]{dynesty}
{Speagle}, J.~S. 2020, \mnras, 493, 3132, \dodoi{10.1093/mnras/staa278}

\bibitem[{{Stappers} {et~al.}(2018){Stappers}, {Keane}, {Kramer}, {Possenti}, \& {Stairs}}]{ska}
{Stappers}, B.~W., {Keane}, E.~F., {Kramer}, M., {Possenti}, A., \& {Stairs}, I.~H. 2018, Philosophical Transactions of the Royal Society of London Series A, 376, 20170293, \dodoi{10.1098/rsta.2017.0293}

\bibitem[{Taylor {et~al.}(2021)Taylor, Baker, Hazboun, Simon, \& Vigeland}]{entex}
Taylor, S.~R., Baker, P.~T., Hazboun, J.~S., Simon, J., \& Vigeland, S.~J. 2021, enterprise\_extensions.
\newblock \url{https://github.com/nanograv/enterprise_extensions}

\bibitem[{Taylor {et~al.}(2024)Taylor, Baker, Hazboun, Simon, Vigeland, {et~al.}}]{entex_edit}
Taylor, S.~R., Baker, P.~T., Hazboun, J.~S., {et~al.} 2024, {mattpitkin/enterprise\_extensions: Keep float128 precision}, float128\_precision,  Zenodo, \dodoi{10.5281/zenodo.13274450}

\bibitem[{{Ushomirsky} {et~al.}(2000){Ushomirsky}, {Cutler}, \& {Bildsten}}]{2000MNRAS.319..902U}
{Ushomirsky}, G., {Cutler}, C., \& {Bildsten}, L. 2000, \mnras, 319, 902, \dodoi{10.1046/j.1365-8711.2000.03938.x}

\bibitem[{{Woan} {et~al.}(2018){Woan}, {Pitkin}, {Haskell}, {Jones}, \& {Lasky}}]{min_ellipticity}
{Woan}, G., {Pitkin}, M.~D., {Haskell}, B., {Jones}, D.~I., \& {Lasky}, P.~D. 2018, \apjl, 863, L40, \dodoi{10.3847/2041-8213/aad86a}

\bibitem[{Xu {et~al.}(2023)Xu, Chen, Guo, Jiang, Wang, Xu, Xue, Caballero, Yuan, Xu, Wang, Hao, Luo, Lee, Han, Jiang, Shen, Wang, Wang, Xu, Wu, Manchester, Qian, Guan, Huang, Sun, \& Zhu}]{Xu_2023}
Xu, H., Chen, S., Guo, Y., {et~al.} 2023, Research in Astronomy and Astrophysics, 23, 075024, \dodoi{10.1088/1674-4527/acdfa5}

\end{thebibliography}
\bibliographystyle{aasjournal}

\end{document}